\documentclass[prl,epsfig,a4paper,onecolumn,noshowpacs,floatfix,groupedaddress,
amsmath,amsfonts,amssymb,preprintnumbers,nofootinbib,citeautoscript]{revtex4}
\usepackage[T1]{fontenc}
\usepackage[latin1]{inputenc}
\usepackage{amssymb,amsmath,bm,dcolumn,graphicx}
\usepackage{color,booktabs,wrapfig,microtype,afterpage} \graphicspath{{Figures/}}
\usepackage[charter,greekuppercase=italicized]{mathdesign}
\usepackage{sidecap}

\makeatletter\renewcommand{\fnum@figure}[1]{\textbf{\sffamily\figurename~\thefigure~|\,}}\makeatother
\makeatletter\renewcommand{\fnum@table}[1]{\textbf{\sffamily\tablename~\thetable~|\,}}\makeatother

\definecolor{NatureBlue}{rgb}{0.012,0.3,0.63}
\definecolor{NiceOrange}{rgb}{0.85,0.42,0.21}
\usepackage[colorlinks,plainpages=false,linkcolor=NatureBlue,urlcolor=NatureBlue,
citecolor=NatureBlue,pdfpagemode=UseNone,pdfstartview=FitBH]{hyperref}

\makeatletter
\newcommand{\bibstyle@supplement}{\bibpunct[, ]{[S}{]}{;}{n}{,}{,}%
    \gdef\bibnumfmt##1{[S##1]}}
\makeatother

\begin{document}

\title{\flushleft\fontsize{16pt}{16pt}\selectfont\sffamily \textcolor{NiceOrange}
{Run-and-pause dynamics of cytoskeletal motor proteins}}

\author{\sffamily Anne E. Hafner$^{1,*}$, Ludger Santen$^1$, Heiko Rieger$^1$, M. Reza Shaebani$^{1,*,\dagger}$\smallskip}

\affiliation{\flushleft
\mbox{\sffamily $^1$\hspace{0.5pt}Department of Theoretical Physics, 
Saarland University, 66041 Saarbr\"ucken, Germany}
\mbox{\sffamily $^*$\hspace{0.5pt}These authors contributed equally to this work.}
\newline
\mbox{\sffamily $^\dagger$\hspace{0.5pt}Corresponding author. shaebani@lusi.uni-sb.de}
\vspace*{3pt}}

\begin{abstract}\citestyle{nature}
\parfillskip=0pt\relax\fontsize{9pt}{11pt}\selectfont\noindent\textbf{
\hspace{-2.5mm}
Cytoskeletal motor proteins are involved in major intracellular 
transport processes which are vital for maintaining appropriate 
cellular function. When attached to cytoskeletal filaments, the 
motor exhibits distinct states of motility: active motion along 
the filaments, and pause phase in which it remains stationary 
for a finite time interval. The transition probabilities between 
motion and pause phases are asymmetric in general, and considerably 
affected by changes in environmental conditions which influences 
the efficiency of cargo delivery to specific targets. By considering 
the motion of individual non-interacting molecular motors on a 
single filament as well as a dynamic filamentous network, we 
present an analytical model for the dynamics of self-propelled 
particles which undergo frequent pause phases. The interplay 
between motor processivity, structural properties of filamentous 
network, and transition probabilities between the two states 
of motility drastically changes the dynamics: multiple 
transitions between different types of anomalous diffusive 
dynamics occur and the crossover time to the asymptotic 
diffusive or ballistic motion varies by several orders of 
magnitude. We map out the phase diagrams in the space of 
transition probabilities, and address the role of initial 
conditions of motion on the resulting dynamics.}
\end{abstract}

\pagestyle{plain}
\makeatletter
\renewcommand{\@oddfoot}{\hfill\bf\scriptsize\textsf{\thepage}}
\renewcommand{\@evenfoot}{\bf\scriptsize\textsf{\thepage}\hfill}
\makeatother

\citestyle{nature}
\maketitle

\makeatletter\immediate\write\@auxout{\string\bibstyle{my-nature}}\makeatother
\renewcommand\bibsection{\section*{\sffamily\bfseries\footnotesize References\vspace{-10pt}\hfill~}}

\noindent
The cellular cytoskeleton is a highly dynamic and complex network 
of cross-linked biopolymers, which carries out essential functions 
in cells such as serving as tracks for motor proteins and adjusting 
the shape and spatial organization of the cell. The cytoskeleton 
enables the cell to efficiently adapt to changes in the environment. 
Microtubules (MTs) and actin filaments constitute the dynamic tracks 
for intracellular transport driven by molecular motors \cite{Chowdhury13,
Appert-Rolland15}. The cytoskeletal filaments are polar due to their 
structural asymmetry, with different spatial organizations. While 
the orientations of actin filaments are rather randomly distributed, 
MTs are usually growing outwards from a  microtubule organizing 
center with their plus ends facing the cell periphery. MTs often 
span long distances, of the order of the cell diameter, whereas 
the mesh size of the actin cortex is of the order of $100\,\text{nm}$ 
\cite{Morone16}.

\smallskip\smallskip\smallskip
\noindent
Three different families of motor proteins are involved in the 
active intracellular transport of organelles and other cargoes: 
kinesins, dyneins, and myosins. Molecular motors of each family 
always move on a specific type of track in a particular direction 
\cite{Schliwa03}; while kinesins and dyneins usually move along 
MTs towards the plus and minus ends, respectively, different 
types of myosins move on actin filaments towards the plus or 
minus directions. Motors can bind to or unbind from the filaments. 
When bind to the filaments, they may perform a number of steps 
or remain stationary. Such binding/unbinding and moving/pausing 
cycles are frequently repeated during the motion of motors along 
the cytoskeleton. The tendency of the motor to continue its motion 
along the filament, called {\em motor processivity}, varies with 
the type of filament and motor \cite{Shiroguchi07,Ali07}, and is 
highly sensitive to the environmental conditions such as the 
presence of specific binding domains or proteins \cite{Okada03,
Culver-Hanlon06,Vershinin07}. Active transport on filaments 
makes the long-distance cargo delivery to specific targets in 
cells feasible, whereas the detached phases are extremely 
inefficient since the cytoplasm is a highly crowded environment 
which slows down the transfer of materials \cite{Hofling13,
Bressloff13}. The combination of active and passive motions 
has been however shown to be beneficial for optimizing the 
first-passage properties \cite{Benichou11}.

\smallskip\smallskip\smallskip
\noindent
The motion of molecular motors involves a high degree of 
complexity due to the dynamics of cytoskeletal filaments, 
binding/unbinding and moving/pausing cycles of motion, 
motor-motor interactions, variations of environmental 
conditions, etc. The dynamics of cargoes is even more 
complicated, as they can be transported by teams of 
motors moving in different directions \cite{Hancock14}. 
The movement of motor proteins can be studied at different 
length and time scales \cite{Bressloff13}. There have 
been theoretical studies focusing on the mechanistic 
details of stepping and the chemomechanical energy 
transduction process \cite{Kolomeisky07,Fisher99,Reimann02}. 
However, when considering longer length and timescales, 
the microscopic details of performing single steps are 
often ignored in the majority of the analytical studies. 
Instead, the dynamics of molecular motors have been 
modeled at the mesoscopic level, via random walk models 
\cite{Muller08,Lipowsky05,Klumpp05} or by solving a set 
of partial differential equations with appropriate 
boundary conditions to consider the transitions between 
different states of motility \cite{Smith01,Jung09,Loverdo08,
Bressloff09}. Stochastic two-state models of motion, 
consisting of altering phases of active and passive dynamics, 
have been widely employed to describe the motion of cytoskeletal 
motor proteins \cite{Pinkoviezky13,Thiel12} and swimming 
bacteria \cite{Elgeti15,Theves13,Taktikos13,Soto14}, and 
locomotive patterns in other biological systems \cite{Hofling13,
Bressloff13,Shaebani16a,Angelani13}. The transition 
probabilities between the two states are generally 
asymmetric and influence the efficiency of cargo delivery 
\cite{Taktikos13}. Interestingly, bacteria are capable of 
adjusting the balance between their running and tumbling 
states in response to the changes in environmental 
conditions. For example, it has been recently observed that 
viscoelasticity of the medium suppresses the tumbling phase 
and enhances the swimming speed of E.\ coli \cite{Patteson15}.  

\smallskip\smallskip\smallskip
\noindent
Here, we adopt a coarse-grained approach to study the dynamics 
of individual non-interacting molecular motors on cytoskeleton, 
which enables us to identify the impact of motor processivity, 
structural properties of the filamentous network, and switching 
frequencies between the two phases of motion on the transport 
properties of motors. To isolate the influence of switching 
frequencies, we first consider the motion along a single 
filament and present a random walk model with two states of 
motility: active motion along the filament and pausing periods. 
The pause state emerges when the motor remains attached but 
stationary on the filament, or if it detaches but stays immobile 
at the detachment point until it attaches again. The latter case 
can be practically observed in laterally confined geometries, 
such as transport along a parallel bundle of microtubules in 
axons \cite{Sheng12,Niescier16,Fu13}. It can also happen when 
the detached motor carries a big vesicle or a protein complex. 
In this case, the diffusion constant in the cytoplasm may be 
even a hundred times smaller than for a single motor \cite{Mastro84,
Arrio-Dupont00}, thus, the big complex remains practically 
immobile at the detachment point. Note that we do not 
consider here the unbinding events which are followed by 
passive diffusion in the cytoplasm. The formalism presented 
in this study can be straightforwardly extended to those 
run-and-tumble motions. We focus on the single (non-interacting) 
particle dynamics in this study. Investigation of the motion 
of a group of interacting motors is beyond the scope of the 
present study. The dynamics is more complicated in such cases 
as, for example, persistent motion in the presence of 
volume-excluded interactions between the particles leads to 
jammed regions or even jamming transition with increasing 
the particle density \cite{Soto14,Fily12,Bialke13,Peruani06}. 
The effects of transition probabilities and volume exclusion 
on the density and traffic of motors along the filament was 
recently studied \cite{Pinkoviezky13}. We assume spontaneous 
transitions between the two states of motility, i.e.\ the model 
is Markovian with constant probabilities. Thus, no acceleration 
or deceleration takes place at a switching event and the transition 
between the two states of motility happens instantly. Moreover, 
the transition probabilities are supposed to be independent. In 
general, the probability $\kappa_w$ of switching from motion to 
pause depends on many factors, such as the applied load on the 
motor, cytoplasmic crowding, and the presence of microtubule-associated 
proteins (MAPs) \cite{Schnitzer00,Trinczek99,Mallik04,Goychuk14,
Lopez93}. Some of these factors may also affect the probability 
$\kappa_m$ of switching from pause to motion, however, there are 
examples (such as the applied load) which do not influence 
$\kappa_m$. Hence, we study the most general case, where the 
transition probabilities $\kappa_w$ and $\kappa_m$ are independent 
of each other. Assuming constant transition probabilities in 
our model results in exponential distributions for the residence 
times in each state, which is in agreement with the experimental 
findings for the distribution of active lifetimes of micron-sized 
beads moving along cytoskeleton \cite{Arcizet08}. We derive exact 
analytical expressions for the temporal evolution of the moments 
of displacement and show that depending on the transition probabilities 
between the two states, the motor can experience crossovers between 
different anomalous diffusive dynamics on different time scales.

\smallskip\smallskip\smallskip
\noindent
In the second part of this work we study the motion of 
individual motor proteins on dynamic cytoskeletal filaments 
to disentangle the combined effects of moving/pausing 
transition probabilities, processivity, and structural properties 
of the underlying network on transport dynamics. We introduce 
a coarse-grained perspective to the problem and consider the 
motion of motors at the level of traveling between network 
junctions rather than individual steps along filaments. Since 
the distribution of directional change contains rich information 
about the particle dynamics \cite{Shaebani14,Sadjadi15,Burov13,Shaebani16c}, 
we characterize the structure of the filamentous network 
by probability distributions $R(\phi)$ for the angle 
$\phi$ between intersecting filaments, and $\mathcal{F}(
\ell)$ for the segment length $\ell$ between neighboring 
intersections. The model consists of two states of 
motility: active motion and waiting at the junctions. 
Such waiting periods at junctions have been experimentally 
observed for transport along cytoskeletal networks 
\cite{Ali07,Ross08,Balint13}. For example, tracking the 
motion of lysosomes on MT networks has revealed that the 
particles experience even long pauses at the nodes of 
the network before that they can either pass through 
it or switch to the intersecting filament \cite{Balint13}. 
In the motion state, the motor either moves processively 
along the previous filament or switches to a new one. 
The cytoskeleton is a dynamic network due to the 
underlying growth and shrinkage of filaments. The 
dynamics even varies depending on the cell type and 
region \cite{Shaebani16b}. Thus, the transport takes 
place on a continuously changing structure, which 
justifies the relevance of our stochastic approach 
as the network structure is implicitly given via the 
probability distributions. Within the proposed 
analytical framework, we prove the existence of different 
regimes of anomalous motion and that the motor may 
experience several crossovers between these regimes, 
as observed in various experiments of motion on 
cytoskeletal filaments \cite{Kulic08,Caspi00,Caspi02,
Bruno09,Salman02}. It is also shown that the crossover 
times between different regimes and the asymptotic 
diffusion coefficient can vary widely depending on 
the key parameters of the model: processivity, network 
structure, and transition probabilities. We address the 
role of initial conditions of motion on the resulting 
dynamics \cite{Saxton96}, and validate the theoretical 
predictions by performing extensive Monte Carlo 
simulations. By studying the evolution of the transport 
properties one can identify, for example, the length- 
and timescales on which the system can maintain a beneficial 
type of diffusive dynamics before a transition to another 
type occurs. This can be, e.g., a crossover from sub- 
or supperdiffusion to another anomalous diffusive transport.
Subdiffusion maintains concentration gradients, thus, it 
is beneficial for a variety of cellular functions which 
depend on the localization of the involved reactants 
\cite{Weigel11,Golding06,Guigas08,Sereshki12}. On the 
contrary, long-distance intracellular transport needs 
to be even faster than the normal diffusion, since an 
efficient delivery of materials to their correct location 
within a cell is crucial for maintaining normal cellular 
function. This is indeed achieved by motor-driven transport 
along cytoskeletal filaments. Our powerful formalism enables 
us to study the influence of the key factors on the formation 
of different types of crossovers and their corresponding length- 
and timescales. Thus, one would be able to predict how far 
a desired type of transport remains efficient when the 
influential parameters vary because of the changes in the 
environment, diseases, etc. The proposed analytical approach 
and the results are also applicable to run-and-tumble motions 
in other biological as well as nonliving systems. 

\vspace{5mm}
\noindent
\textbf{\Large Results}

\vspace{2mm}
\noindent
\textbf{\large Motion along a single filament} 

\begin{figure}[t]
\centering
\includegraphics[width=\textwidth]{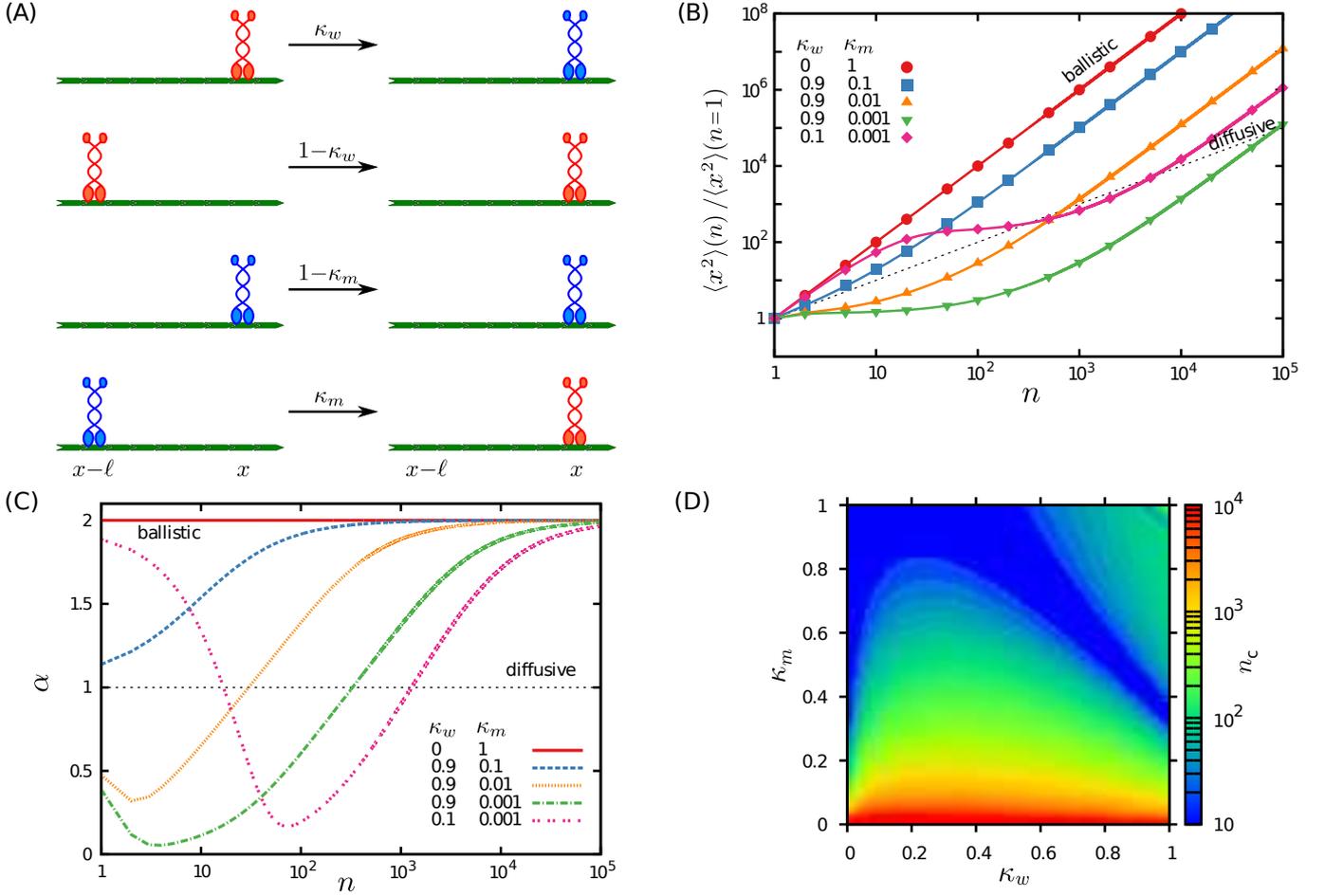}
\caption{{\bf Motion on a single filament.} (A) A 
schematic of the transition probabilities between 
pause (blue) and motion (red) states of a motor protein 
along a single cytoskeletal filament, as described by the 
set of master equations (\ref{Eq:MasterEqs-1D}). The four 
possibilities for the motility states of two successive 
steps are shown separately. (B) MSD as a function of the 
step number $n$ for $\lambda{=}1$, $P_0^M{=}1$, and 
several values of $\kappa_m$ and $\kappa_w$. The solid 
lines are obtained from the analytic expression (S17) 
and the symbols represent Monte Carlo simulation results. 
(C) Temporal evolution of the anomalous exponent $\alpha$ 
via Eq.\,(\ref{Eq:Exp-1D}), for the same parameters 
as in panel (B). (D) The crossover time $n_c$ to the 
asymptotic ballistic motion for $\lambda{=}1$ and 
$P_0^M{=}1$ in the ($\kappa_w$, $\kappa_m$) phase space.}
\label{Fig:1}
\end{figure}

\vspace{2mm}
\noindent
We first consider the motion of a molecular motor on a single 
filament in a crowded environment. For spontaneously switching 
motors, the stochastic motion can be described by two states 
of motility: (i) \emph{ballistic motion} along the filament, 
and (ii) \emph{pause} phase, where the motor remains 
stationary along the filament for a while (or unbinds from 
the filament but stays immobile until binds again). When 
the motor restarts the active motion, it continues 
towards its previous direction (unidirectional motion). 
The switching from motion to pause phase and vice 
versa happens with probabilities $\kappa_w$ and $\kappa_m$, 
respectively. The case of considerable displacements during 
the detached periods is not considered in the present study. 
The transition probabilities between the two states of motion 
are not necessarily symmetric in general, thus, we consider 
asymmetric constant transition probabilities [see Fig.~\ref{Fig:1}(A)]. 
The assumption of constant transition probabilities leads 
to exponential probability distributions $P_\text{waiting}(t)$ 
and $P_\text{motion}(t)$ for the residence times in the waiting 
and motion states, respectively. For example, one can write 
the master equation $P_\text{waiting}(t)=P_\text{waiting}(t
{-}1)\,(1{-}\kappa_m)$ in the waiting state, and solve it 
recursively to obtain $P_\text{waiting}(t) = \frac{\kappa_m}{1
{-}\kappa_m}\,\text{e}^{\ln(1{-}\kappa_m)\,t}$. The decay 
exponent of the exponential distribution is proportional 
to $\ln(1{-}\kappa_m)$ and $\ln(1{-}\kappa_w)$ for the 
waiting and motion states, respectively. Thus, a smaller 
transition probability from state I to II, leads to a 
slower exponential decay for the residence time distribution 
in state I, resulting in a longer average lifetime in this 
state. At each time step, the particle either waits or 
performs a step of length $\ell$ taken from a probability 
distribution $\mathcal{F}(\ell)$. We introduce the probability 
densities $P_{n}^{M}(x)$ and $P_{n}^{W}(x)$ to find the 
walker at position $x$ at time step $n$ in the motion 
and waiting states, respectively. The temporal evolution 
of the process can be described by the following set 
of coupled master equations
\begin{equation}
\left\{ 
\begin{array}{ll}
P_{n+1}^M(x) = \displaystyle\int \!\!\!\! \text{d}\ell \, 
\mathcal{F}(\ell) \Big[\kappa_mP_n^W(x{-}\ell) 
{+} (1-\kappa_w)P_n^M(x{-}\ell)\Big],\\
\\
P_{n+1}^W(x)=\kappa_w P_n^M(x){+}(1{-}\kappa_m)P_n^W(x).
\end{array}
\right.
\label{Eq:MasterEqs-1D}
\end{equation}
We develop a Fourier-z-transform approach \cite{Sadjadi08,
Sadjadi11,Miri05}, which enables us to obtain exact 
analytical expressions for the arbitrary moments of 
displacement. The details of the theoretical approach 
can be found in the {\em Suppl Info}, where as an 
example, the lengthy expression for the mean squared 
displacement (MSD) is obtained. It can be seen from 
Eq.(S17) that, in addition to the transition probabilities, 
the results also depend on the initial conditions of 
motion. In derivation of Eq.(S17), denoting the 
probability of initially starting in the motion state 
by $P_0^M$, the following initial conditions are 
imposed 
\begin{align}
&P_{n{=}0}^M(x) = P_0^M \, \delta(x),\notag \\
&P_{n{=}0}^W(x) = (1-P_0^M) \, \delta(x),
\label{Eq:InitCond-1D}
\end{align}
which results in $P_{n{=}0}(x) {=} P_{n{=}0}^M(x) {+} 
P_{n{=}0}^W(x) {=} \delta(x)$. The analytical results 
for the time evolution of MSD are shown in 
Fig.~\ref{Fig:1}(B) for different values of $\kappa_m$ 
and $\kappa_w$ parameters. The analytical predictions 
are also validated by performing extensive Monte Carlo 
simulations. The quantities of interest are obtained 
at discrete times by ensemble averaging over $10^6$ 
realizations. More details on the simulation method 
can be found in the \emph{Suppl Info}. Several 
crossovers on different time scales can be observed, 
even though the asymptotic behavior is ballistic 
in all cases as expected from the unidirectional 
motion of the motor along the filament. In order 
to determine the crossovers more quantitatively, 
we fit the time dependence of the MSD to a power-law 
$\langle x^2 \rangle {\sim} t^\alpha$, with $\alpha$ 
being the anomalous exponent. For instance, the 
following expression can be deduced for the 
initial anomalous exponent by fitting the first 
two steps of motion
\begin{align}
\alpha^\star = \ln\left[2 - \frac{2(\kappa_w{-}
1)}{\lambda} - \kappa_w + \kappa_m \Big(-1 {+} 
\frac{1}{\kappa_m {-} P_0^M (-1 {+} \kappa_m {+} 
\kappa_w)}\Big)\right]\bigg/\ln[2],
\label{Eq:InitExp-1D}
\end{align}
with $\lambda{=}\langle \ell^2\rangle /\langle 
\ell \rangle^2$ being the variance of the step-length 
distribution. Assuming a constant step length for 
simplicity (i.e.\ $\lambda{=}1$), if the walker 
remains in the motion phase forever (i.e.\ 
$\kappa_w{=}0$ and $\kappa_m{=}1$) one gets 
$\alpha^\star {=} 2$ as expected for a ballistic 
motion. When starting from the motion state, one 
has $P_{n{=}0}^W(x){=}0$ and $P_{n{=}0}^M(x) {=} 
\delta(x)$, and the initial slope follows  
\begin{align}
\alpha^\star = \ln\left[4 - 3 \kappa_w - \kappa_m 
+ \frac{\kappa_m}{1-\kappa_w} \right] \bigg/ \ln[2].
\label{Eq:InitExp-1D-Motion}
\end{align} 

\noindent
By applying the fit to successive time steps $n$ 
and $n{+}1$, the anomalous exponent as a function 
of time can be obtained as
\begin{align}
\alpha(n) =\ln \left[\frac{\langle r^2 \rangle_{n+1}
}{\langle r^2 \rangle_n}\right] \bigg/ \ln 
\left[\frac{n+1}{n}\right].
\label{Eq:Exp-1D}
\end{align}
Figure~\ref{Fig:1}(C) shows the time evolution of the 
exponent for a motor starting from the motion state. 
The slope $\alpha$ of the MSD curve is influenced 
by: (i) the initial conditions of motion $P_0^M$ and 
$P_0^W{=}1{-}P_0^M$, (ii) the transition probabilities 
$\kappa_m$ and $\kappa_w$ which determine the stationary 
probabilities $P_\infty^M$ and $P_\infty^W$ of being at each 
state (see Eq.(15) of the discussion section), and (iii) 
the asymptotic dynamics. If the initial probabilities $P_0^M$ 
and $P_0^W$ are different from the stationary ones, a transition 
to the stationary probabilities $P_\infty^M$ and $P_\infty^W$ 
occurs, thus, $\alpha$ undergoes a crossover at short times. 
Another crossover happens at longer times towards the asymptotic 
dynamics. In the examples shown in Fig.1(C), the initial conditions 
are chosen to be $P_0^W{=}0$ and $P_0^M{=}1$, i.e.\ the walk always 
starts in the motion state. Let us consider the case of $\kappa_w
{=}0.1$ and $\kappa_m{=}0.001$, as an example. The convergence 
to the stationary probabilities happens after less than 100 
steps for this set of parameters (see the discussion section 
for details). Therefore, the first change in the 
curvature of $\alpha$ vs $n$ in Fig.1(C) at short times is due 
to the crossover from initial to stationary conditions. Starting 
from motion with $\kappa_w{=}0.1$, it takes 10 steps, on average, 
that a transition to the waiting state happens. That is why the 
initial slope is supperdiffusive ($1{<}\alpha$). Then, because 
of the very low transition probability $\kappa_m{=}0.001$, the 
motor typically stays in the waiting state for long times which 
results in a subdiffusive dynamics ($\alpha{<}1$). Finally, the 
second change of the curvature of $\alpha$ at longer times 
($n{\sim}1000$) evidences the crossover to the asymptotic ballistic 
motion ($\alpha{=}2$) induced by the unidirectionality of the motion.

\smallskip\smallskip\smallskip
\noindent
In the limit of $n {\to} \infty$ the terms proportional 
to $n^2$ dominate and the motility becomes purely 
ballistic. The asymptotic MSD is given by
\begin{align}
\langle x^2 \rangle_{n\rightarrow\infty} \approx 
\frac{\kappa_m^2 (\kappa_m{+}1)^2}{(\kappa_m{+}
\kappa_w)^4} \, \langle\ell\rangle^2 \, n^2.
\label{Eq:MSDasymp-1D}
\end{align}
The prefactor depends on the switching probabilities. If 
the walker never switches to the waiting state, one 
recovers the relation $\langle x^2 \rangle {=} \langle 
\ell \rangle^2 n^2$ which is the fastest possible 
propagation. The crossover to the long-term ballistic 
regime is approached asymptotically. We use the 
distance from the exponent of ballistic motion, i.e.\ 
$\delta\alpha =|\alpha(n){-}2|$, and estimate the 
crossover time as the time step $n_c$ at which 
$\delta\alpha$ drops below a threshold value $\epsilon$. 
Here we report the results for $\epsilon{=}10^{-2}$, 
however, we checked that the choice of $\epsilon$ 
does not affect our conclusions. As it can be seen 
from Fig.~\ref{Fig:1}(D), $n_c$ varies by several 
orders of magnitude in the ($\kappa_w$, $\kappa_m$) 
plane. It is expected that $n_c$ increases with 
decreasing $\kappa_m$, as the chance of switching 
to the motion state is reduced. The increase of 
$n_c$ at large values of both probabilities 
$\kappa_m$ and $\kappa_w$ is due to frequent state 
oscillations which postpone the transition to 
asymptotic ballistic regime to longer times. We note 
that the length scale over which the motor reaches 
the asymptotic ballistic motion along a single 
filament is usually smaller or comparable to the 
cell size. Assuming that a motor protein moves with 
steps of size ${\sim}8\,\text{nm}$ and a typical 
velocity $v {\sim} 1\,\mu\text{m}{/}\text{s}$ 
\cite{Ali08}, its displacement (until the transition 
to the asymptotic dynamics happens) ranges from $0.1$ 
to $100\,\mu\text{m}$, which is comparable to typical 
cell sizes. Thus, the motor can practically experience 
the transition to long-term ballistic motion along a 
single microtubule within the cell body for a 
considerable range of the transition probabilities 
$\kappa_w$ and $\kappa_m$.

\vspace{5mm}
\noindent
\textbf{\large Motion on a dynamic filamentous network} 

\vspace{2mm}
\noindent
The theoretical framework can be generalized to describe 
active motion of particles on filamentous networks. We 
adopt a coarse-grained approach in which the motion of 
motor proteins is modeled as a persistent random walk 
on the intersections of cytoskeletal filaments. The 
structure of the network is characterized by the probability 
distributions $R(\phi)$ for the angle $\phi$ between 
intersecting filaments, and $\mathcal{F}(\ell)$ for the 
segment length $\ell$ between neighboring intersections. 
Furthermore, a parameter $p$ is introduced to take the 
processivity of molecular motors into account. The particle 
either waits at each time step (waiting state) or walks with 
a step length $\ell$ (motion state). In the latter state, 
the motor either continues along the previous filament with 
probability $p$ or chooses a new filament with probability 
$1{-}p$. Similar to motion on a single filament, the transition 
probabilities $\kappa_m$ and $\kappa_w$ between the two states 
are assumed to be asymmetric and constant. The parameter 
$\kappa_m$ ($\kappa_w$) denotes the switching probability 
from waiting to motion (motion to waiting) state. The 
probability density functions $P_{n}^{M}(x,y|\theta)$ 
and $P_{n}^{W}(x,y|\theta)$ denote the probability to 
find the walker at position $(x,y)$ along the direction 
$\theta$ at time step $n$ in the motion and waiting states, 
respectively. Here a 2D system is considered for simplicity. 
For extension of the approach to 3D see \cite{Sadjadi15}. 
The evolution of the process is described by 
the following set of coupled master equations
\begin{equation}
\left\{ \begin{array}{ll}
P_{n+1}^{M}(x,y|\theta) = \\ 
p \!\!\displaystyle\int\!\!\! d\ell \, \mathcal{F}(\ell) 
\, \Bigg[\!\kappa_m P^{W}_{n}\!\Big(x'\!,y'\big|\theta
\Big) {+} (1{-} \kappa_w) \, P^{M}_{n} \! \Big(x'\!,y'
\big|\theta\Big)\!\Bigg] {+} (1{-}p) \!\! \int \!\!\! 
d\ell \, \mathcal{F}(\ell) \! \int_{-\pi}^{\pi} 
\!\!\!\!\!\!\!\! d\gamma \, R(\phi) \, \Bigg[\!\kappa_m 
P^{W}_{n}\!\Big(x'\!,y'\big|\gamma\Big) {+} (1{-} 
\kappa_w) \, P^{M}_{n} \! \Big(x'\!,y'\big|\gamma
\Big)\!\Bigg],\vspace{2mm}\\
\\
P_{n+1}^{W}(x,y|\theta) = \kappa_w P_{n}^{M}\!
(x,y|\theta){+}(1{-}\kappa_m) P_{n}^{W}\!(x,y|\theta).
\end{array}
\right.
\label{Eq:MasterEqs-2D}
\end{equation}

\begin{figure}
\centering
\includegraphics[width=\textwidth]{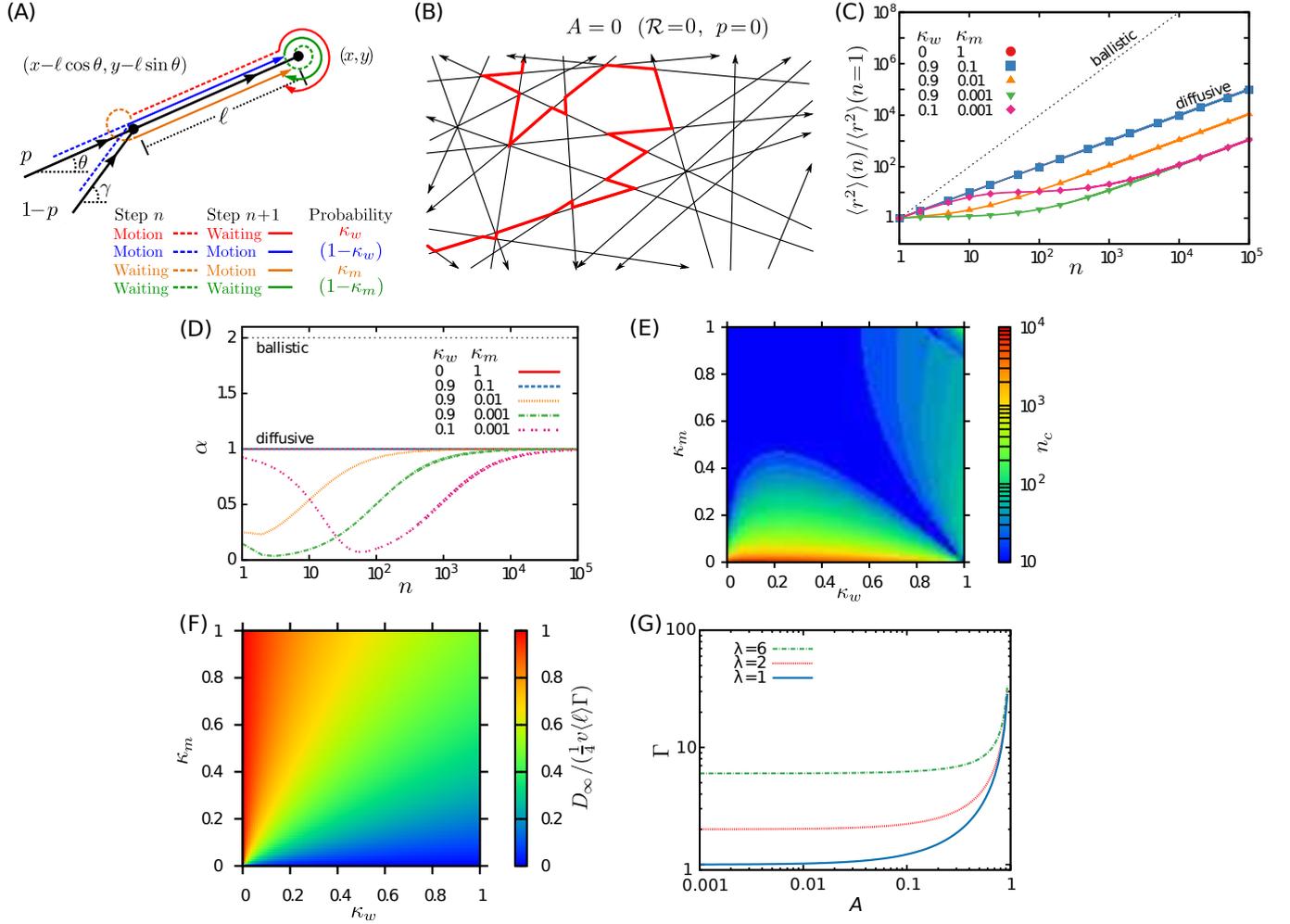}
\caption{{\bf Non-processive motion on random actin networks.} 
(A) Schematic of the path on a cytoskeletal network 
during two consecutive steps of motion, as described by 
the coupled set of master equations (\ref{Eq:MasterEqs-2D}). 
(B) A typical sample trajectory (red lines) of a 
non-processive walker ($p{=}0$) on a random filamentous 
structure ($\mathcal{R}{=}0$). (C) MSD as a function 
of the step number $n$ for $\lambda{=}1$, $P_0^M{=}1$, 
$A{=}0$, and several values of $\kappa_m$ and $\kappa_w$. 
The solid lines correspond to the analytical expression 
(S45) and the symbols represent the simulation results. 
(D) Temporal evolution of the anomalous exponent 
$\alpha$ via Eq.\,(\ref{Eq:Exp-1D}), for the same 
parameters as in panel (C). (E,F) Phase diagrams 
of (E) the crossover time $n_c$ to the asymptotic 
diffusive regime, and (F) the long-term diffusion 
constant $D_\infty$, scaled by $\frac14 v \langle 
\ell \rangle \Gamma$, via Eq.\,(\ref{Eq:Dinfty}) 
for $A{=}0$, $\lambda{=}1$, and $P_0^M{=}1$ in the 
($\kappa_w$, $\kappa_m$) plane. (G) The scale 
parameter $\Gamma{=}\frac{A(\lambda-2)-\lambda}{A-1}$ 
vs $A$ for several values of $\lambda$.}
\label{Fig:2}
\end{figure}
\begin{figure}
\centering
\includegraphics[width=\textwidth]{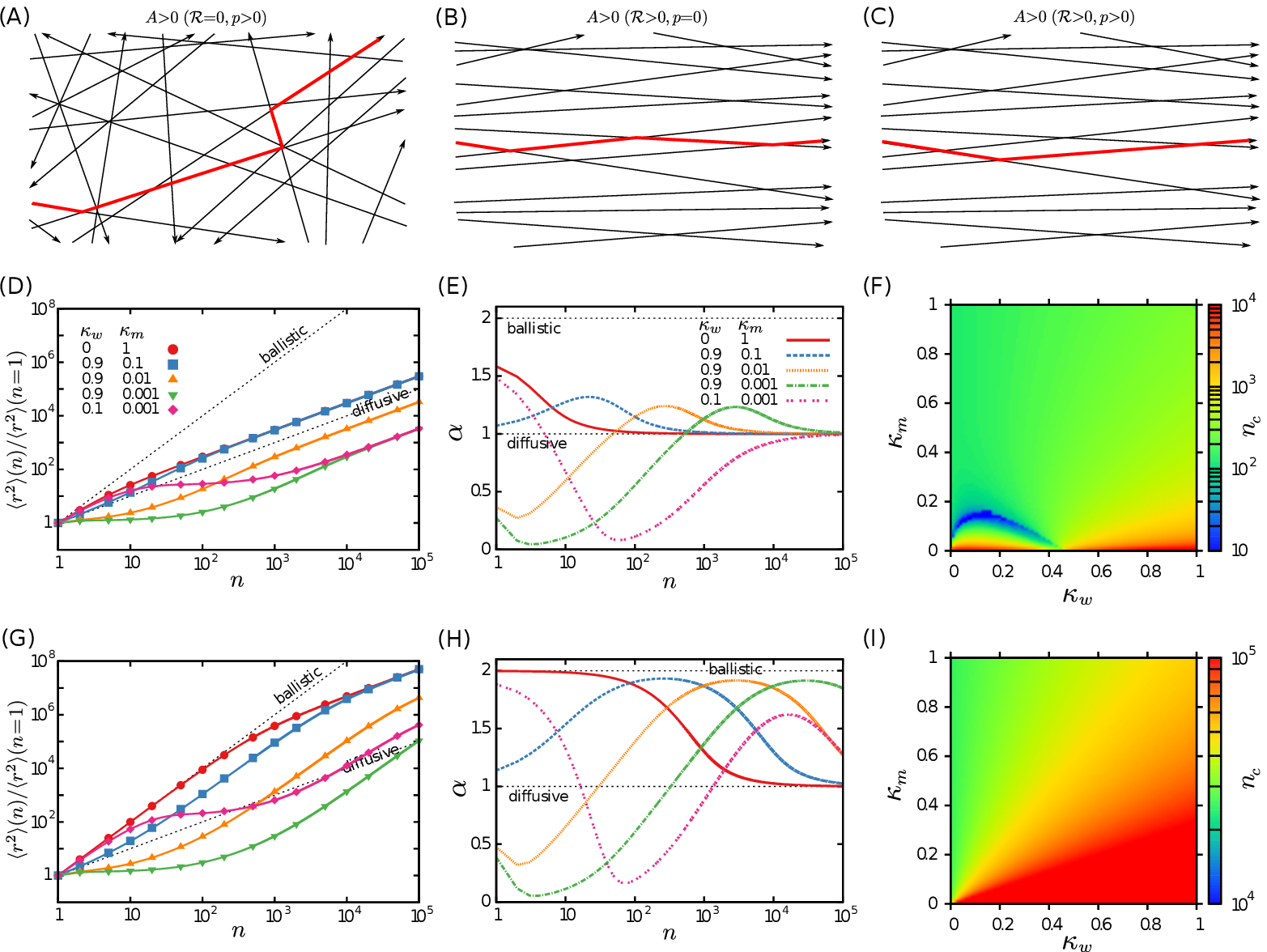}
\caption{{\bf Motion on microtubule networks / 
Processive motion.} In the upper panels, typical 
sample trajectories (red lines) are shown for 
positive values of the parameter $A$: (A) a 
processive motion ($p{>}0$) on a random filamentous 
structure ($\mathcal{R}{=}0$), (B) a non-processive 
motion ($p{=}0$) on a bundle of relatively parallel 
filaments ($\mathcal{R}{>}0$), and (C) a processive 
motion on a bundle of relatively parallel filaments 
(which corresponds to an extremely high value of 
$A$). In the middle and lower panels, the results 
for $A{=}0.5$ and $0.99$ are shown, respectively. 
The latter value is relevant for a processive motion 
on highly aligned filaments, while the former case 
corresponds to either a non-processive motion on 
moderately aligned filaments or, alternatively, 
active motion with moderate processivity on 
isotropic networks. (D,G) MSD in terms of the 
step number $n$ for $\lambda{=}1$, $P_0^M{=}1$, 
and several values of $\kappa_m$ and $\kappa_w$. 
The solid lines (symbols) correspond to analytical 
results via Eq.(S45) (simulation results). (E,H) 
Time evolution of the anomalous exponent $\alpha$ 
for the same parameter values as in panels 
(D,G). (F,I) Crossover time $n_c$ to the 
asymptotic diffusive dynamics in the case 
of $\lambda{=}1$ and $P_0^M{=}1$ in the 
($\kappa_w$, $\kappa_m$) phase space.}
\label{Fig:3}
\end{figure}

\noindent
Assuming isotropic initial conditions
\begin{align}
P_0^M(x,y|\theta)&=\frac{1}{2\pi}\,
\delta(x)\delta(y)\,P_0^M,\notag\\
P_0^W(x,y|\theta)&=\frac{1}{2\pi}\,
\delta(x)\delta(y)\,(1-P_0^M),
\label{Eq:InitCond-2D}
\end{align}
which leads to $P_0(x,y|\theta)=P_0^M(x,y|\theta)
+P_0^W(x,y|\theta)=\frac{1}{2\pi}\,\delta(x)
\delta(y)$, one can follow the proposed Fourier-z-transform 
formalism and calculate arbitrary moments of 
displacement. The exact analytical expression 
for the MSD, which is the main quantity of 
interest, is given in Eq.(S45). The interplay 
between transition probabilities, motor 
processivity, initial conditions of motion, 
and structure of the underlying network lead 
to a variety of anomalous transport dynamics 
on different time scales. The structure of the 
network is characterized by the relative variance 
of $\mathcal{F}(\ell)$ (i.e.\ $\lambda{=}\langle 
\ell^2\rangle /\langle \ell \rangle^2$) and the 
Fourier transform $\mathcal{R}$ of $R(\phi)$, which 
quantify the heterogeneity and anisotropy of 
the environment, respectively. $\mathcal{R}$ 
ranges from $-1$ to $1$, with $\mathcal{R}{=}0$ 
for a completely random structure, and positive 
(negative) values for an increased probability 
for motion in the near forward (backward) 
directions. The processivity $p$ and anisotropy 
$\mathcal{R}$ parameters always appear combined 
as $A=p+\mathcal{R}-p\mathcal{R}$ in the solution; 
thus, by varying $A$ one can separately study 
the effects of $p$ or $\mathcal{R}$ on the 
transport of motors. We first choose $A{=}0$ in 
Fig.~\ref{Fig:2}, corresponding to a non-processive 
motion ($p{=}0$) on an isotropic structure such as 
actin filament networks ($\mathcal{R}{=}0$), and then 
study positive values $A{=}0.5$ and $A{=}0.99$ in 
Fig.~\ref{Fig:3}. These latter choices can e.g.\ 
correspond to either a non-processive motion ($p{=}0$) 
on aligned filaments ($\mathcal{R}{>}0$) such as 
radially organized MT networks, or an active motion 
($p{>}0$) on an isotropic actin network ($\mathcal{R}
{=}0$). 

\smallskip\smallskip\smallskip
\noindent
As the phase space of the system is entangled, 
we restrict ourselves to initially starting from 
the motion state ($P_0^M{=}1$) in this section, 
and elaborate on the role of initial conditions 
in the discussion section. A wide range of different 
types of anomalous motion can be observed on varying 
the transition probabilities $\kappa_w$ and $\kappa_m$. 
The possible crossovers at short and intermediate 
time scales are even more diverse than those 
observed for the motion on a single filament. 
A similar approach as in the previous section is 
followed to obtain the time evolution of the 
anomalous exponent and identify the crossovers. 
The long term dynamics is diffusive in all cases 
as the directional memory is short ranged and 
the walker eventually gets randomized on the 
network. However, we note that the asymptotic 
diffusive motion might be observable only on 
time and length scales which are not accessible 
in experiments. By fitting to the power-law  
$\langle x^2 \rangle {\sim} t^\alpha$, the 
initial anomalous exponent is obtained as
\begin{align}
\alpha^\star = \ln\left[2 - \frac{2(p{+}
\mathcal{R}{-}p\,\mathcal{R})(\kappa_w{-}
1)}{\lambda} - \kappa_w + \kappa_m \Big(-1 
{+} \frac{1}{\kappa_m {-} P_0^M (-1 {+} 
\kappa_m {+} \kappa_w)}\Big)\right]\bigg
/\ln[2],
\label{Eq:InitExp-2D}
\end{align}
which in the case of $\kappa_w{=}0$ and 
$\kappa_m{=}1$ reduces to 
\begin{align}
\alpha^\star = 1+ \ln\left[1+\frac{p{+}
\mathcal{R}{-}p\,\mathcal{R}}{\lambda}
\right]\bigg/\ln[2].
\label{Eq:InitExp-2D-motion}
\end{align}
For example, one gets $\alpha^\star{=}1$ 
for non-processive motion on random actin networks 
in the absence of waiting phases, which is 
expected to be diffusive on all time scales.  
From Eq.\,(\ref{Eq:InitExp-2D}) one can also 
see the impact of the initial conditions of 
motion and heterogeneity of the network on 
the exponent. 

\smallskip\smallskip\smallskip
\noindent
In the long-time limit, the terms linear in 
$n$ dominate, thus, the motility becomes 
purely diffusive. From Eq.(S45), the MSD in 
the limit of $n {\to} \infty$ is given by
\begin{align}
\langle r^2 \rangle_{n\rightarrow\infty} 
\approx \frac{\kappa_m}{\kappa_m+\kappa_w}\, 
\frac{A(\lambda-2)-\lambda}{A-1}\,\langle
\ell \rangle^2 \, n,
\end{align}
from which the asymptotic diffusion 
constant can be determined as
\begin{align}
D_\infty = \frac14 v\langle\ell\rangle 
\frac{\kappa_m}{\kappa_m+\kappa_w}\, 
\frac{A(\lambda-2)-\lambda}{A-1} = 
\frac14 v\langle\ell\rangle \Gamma \, 
\frac{\kappa_m}{\kappa_m+\kappa_w},
\label{Eq:Dinfty}
\end{align}
with $\Gamma{=}\frac{A(\lambda-2)-
\lambda}{A-1}$ and $v$ being the average motor 
velocity. Figure~\ref{Fig:2}(F) shows how 
$D_\infty$ varies in the space of transition 
probabilities. While decreasing $\kappa_w$ or 
increasing $\kappa_m$ enhances the diffusion 
coefficient, changing the probabilities in the opposite 
direction leads to a strong localization. Taking 
the dependence of $\Gamma$ on the heterogeneity 
parameter $\lambda$ into account [see 
Fig.~\ref{Fig:2}(G)], it can be seen that $D_\infty$ 
varies by several orders of magnitude by varying 
the key parameters of the problem. Interestingly, 
the long-term diffusion constant does not depend 
on the initial conditions. 

\begin{figure*}
\centering
\includegraphics[width=0.7\textwidth]{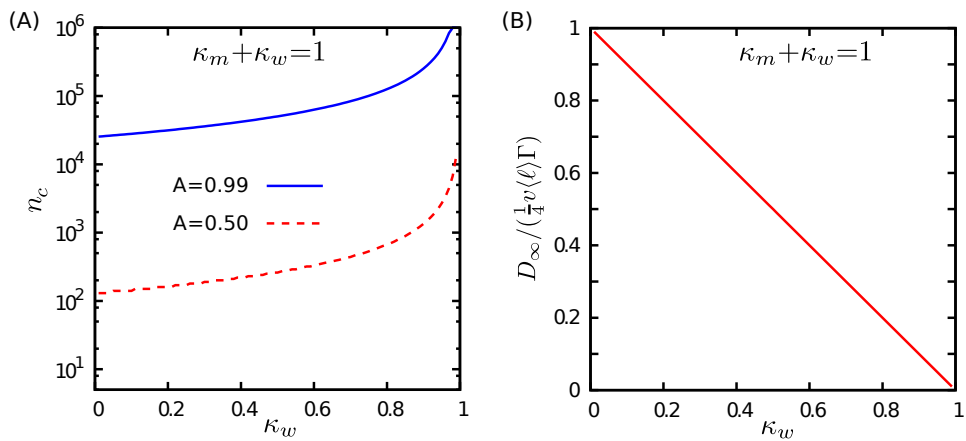}
\caption{(A) The crossover time $n_c$ to the asymptotic 
diffusive dynamics for different values of parameter 
$A$, and (B) the asymptotic diffusion constant $D_\infty$ 
in terms of $\kappa_w$, when the transition probabilities 
are constrained to $\kappa_m{+}\kappa_w{=}1$.}
\label{Fig:4}
\end{figure*} 

\smallskip\smallskip\smallskip
\noindent
To estimate the transition time $n_c$ to asymptotic 
diffusive regime, we follow the convergence of the 
anomalous exponent towards $1$, and determine $n_c$ 
as the earliest time step at which the exponent 
difference drops below a threshold $\epsilon$, i.e.\ 
\begin{align}
\left|\alpha(n) -1 \right| < \epsilon.
\label{eq:nc_network}
\end{align}
The results shown in Figs.~\ref{Fig:2}(E), \ref{Fig:3}(F), 
and \ref{Fig:3}(I), reported for $\epsilon{=}10^{-2}$, 
reveal that $n_c$ varies over several orders of 
magnitude. On an actin filament network with an 
average mesh size $\langle \ell \rangle {\sim} 
100 \text{nm}$ and a typical motor velocity $v
{\sim}1\mu\text{m}{/}\text{s}$, $n_c$ ranges from 
$0.1$ seconds to more than $100$ seconds, which might 
be beyond the possible time window of experiments, 
thus, all different regimes of motion are not 
necessarily realized in practice. It is difficult 
to make a direct comparison with the reported 
experiments due to the lack of the required data. 
Still, let us consider the motion of a motor protein 
with steps of size $8\,\text{nm}$ and a typical 
velocity $v {\sim} 1\,\mu\text{m}{/}\text{s}$ along 
microtubules, as an example. Starting from the motion 
state, if we adjust the transition probabilities to 
$\kappa_w{=}0.9$ and $\kappa_m{=}0.01$ and the 
processivity to $A{=}0.99$, our formalism predicts 
a crossover at short times (${\sim} 100 \text{ms}$) 
from sub- (with $\alpha{\simeq}0.6$) to superdiffusion 
(with $\alpha{\simeq}1.4$), which is quantitatively 
comparable to the transition reported in experiments 
on the bidirectional organelle transport along microtubules 
\cite{Kulic08}. To compare with other analytical and 
numerical studies,  we first adjust our model parameters 
to those of a single-state (i.e.\ $\kappa_w{=}0$ and 
$\kappa_m{=}1$) active motion along a square lattice 
($\mathcal{R}{=}\frac13$) and compare the initial anomalous 
exponents with those obtained from a random velocity 
model (RVM) for active transport on a similar structure 
\cite{Kahana08}. We get initial slopes ranging between 
$1.41$ and $2$ depending on the processivity of the walker, 
which are in agreement with the initial slopes shown in 
Fig.(6) of the RVM paper (ranging within $[4{/}3,2]$ 
upon varying the processivity). However, there are major 
differences between the two models: (i) The disorder is 
quenched in the RVMs, in contrast to the dynamic random 
networks in our model, and (ii) the combination of the 
stochastic processes in the RVMs results in very different 
anomalous transport on intermediate and long timescales 
and the asymptotic dynamics is not a normal diffusion 
necessarily. As a final comparison, an anomalous exponent 
$\alpha^\star{=}1.44$ was recently reported in simulations 
where the particle experiences altering phases of active 
motion along random filaments and passive diffusion in the 
cytoplasm \cite{Ando15}. The displacements of the particle 
are considerably large in the latter phase, of the order 
of those in the active run phase. We can approximate this 
motion with a single-state non-processive motion ($\kappa_w
{=}0$, $\kappa_m{=}1$, $p{=}0$) in our model, where the 
turning-angle distribution of the particle is a combination 
of a uniform distribution (for the passive diffusion phase) 
and a delta function in the forward direction (for the run 
phase along filaments). This leads to a quantitatively 
comparable effective initial exponent $\alpha^\star{\simeq}1.5$.

\smallskip\smallskip\smallskip
\noindent
We note that the assumption of independent transition 
probabilities $\kappa_w$ and $\kappa_m$ enables us to study 
the most general case and obtain the quantities of interest 
in the ($\kappa_m$, $\kappa_w$) phase space. Indeed, this phase 
diagram contains the results for any functionality between 
$\kappa_w$ and $\kappa_m$. In some cases, the influential 
factors affect the transition probabilities in opposite 
directions. For example, increasing the density of MAPs 
on one hand enhances the steric inhibition of the motor 
motility (thus increases $\kappa_w$) and on the other hand 
decreases the chance of restarting the motion (i.e.\ decreases 
$\kappa_m$). In the absence of quantitative experimental studies, 
here we consider the choice $\kappa_m{+}\kappa_w{=}1$ as a 
particular case of inversely related transition probabilities; 
the increase of one of them is accompanied by the decrease of 
the other one. The particular consequence of choosing $\kappa_m
{+}\kappa_w{=}1$ is that the system immediately undergoes 
the crossover to the equilibrated state probabilities 
$P^M_{\!\infty}{=}\kappa_m$ and $P^W_{\!\infty}{=}\kappa_w$, 
as can be seen from Eq.\,(\ref{Eq:MarkovP}) in the next section. 
Therefore, the initial anomalous exponent $\alpha^\star$ does not 
depend on the choice of the initial state probabilities $P^M_0$ 
and $P^W_0$: From Eq.\,(\ref{Eq:InitExp-2D}), one obtains 
$\alpha^\star {=} \ln\left[2 - \frac{2\,A\,(\kappa_w{-}1)}{
\lambda}\right]/\ln[2]$. Moreover, the asymptotic diffusion 
constant via Eq.\,(\ref{Eq:Dinfty}) can be expressed in terms 
of just one of the transition probabilities, e.g.\ $D_\infty 
{=} \frac14 v\langle\ell\rangle \Gamma \,\kappa_m$. Figure 
\ref{Fig:4} shows how $D_\infty$ and the crossover time to 
the asymptotic diffusion vary in terms of the transition 
probabilities when they are constrained to $\kappa_m{+}
\kappa_w{=}1$. The line $\kappa_m{+}\kappa_w{=}1$ divides 
the phase diagram into two subdomains: In the region below 
(above) this line, one or both of $\kappa_w$ and $\kappa_m$ 
can be rather small (large). As we explained in the previous 
section, smaller (larger) transition probabilities lead to 
slower (faster) exponential decays for the residence time 
distributions, resulting in longer (shorter) average 
lifetimes for each state of motility.

\begin{figure*}[t]
\centering
\includegraphics[width=\textwidth]{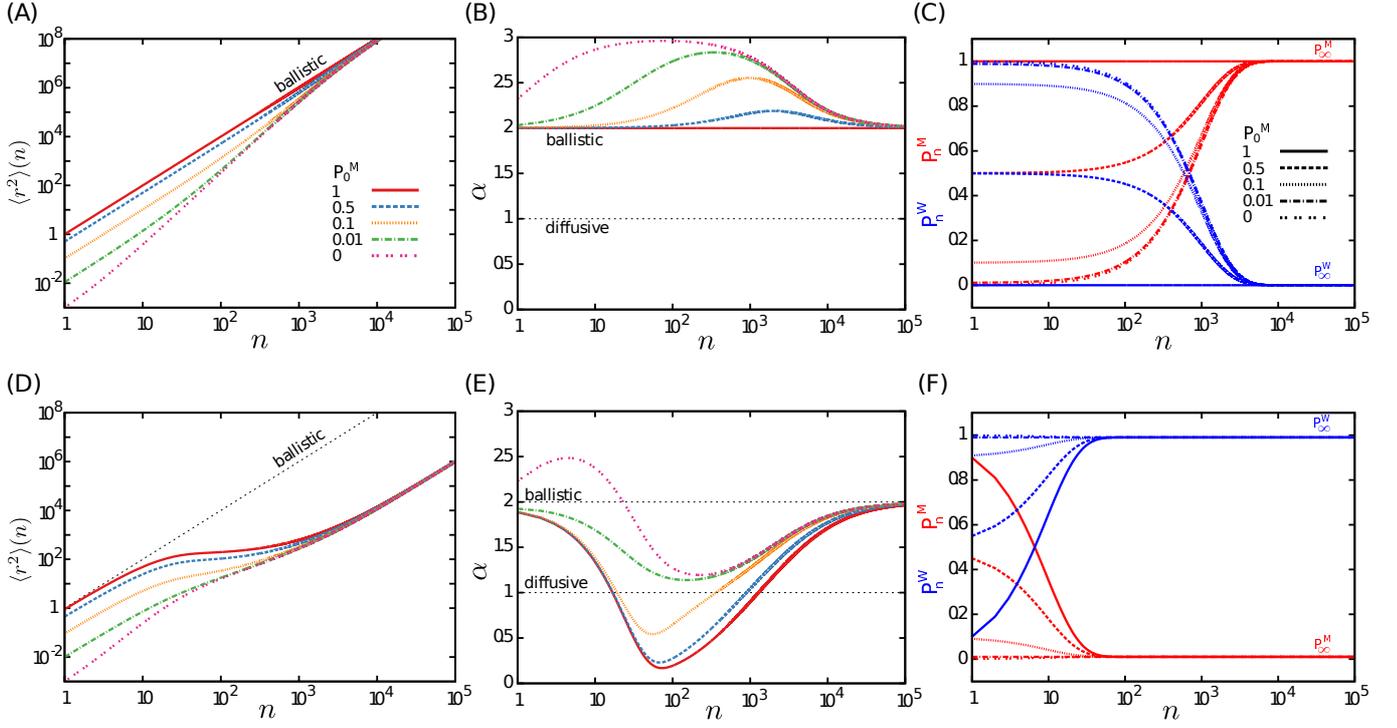}
\caption{{\bf Influence of initial conditions.} 
The dynamics on a single filament in the absence 
of waiting ($\kappa_w{=}0$, upper panels) and 
for weak switching probabilities to waiting state 
($\kappa_w{=}0.1$, lower panels) are compared. 
(A,D) MSD as a function of the step number $n$ 
for $\kappa_m{=}0.001$, $\lambda{=}1$, and 
different probabilities $P_0^M$ of initially 
starting in the motion state. (B,E) Temporal 
evolution of the anomalous exponent $\alpha(n)$ 
for the same set of parameters as in panels (A,D). 
(C,F) Evolution of the Markov chain probabilities 
$P_n^M$ (red lines) and $P_n^W$ (blue lines), via 
Eqs.\,(\ref{Eq:MarkovP}). The stationary probabilities 
are given by Eqs.\,(\ref{Eq:MarkovPstationary}).}
\label{Fig:5}
\end{figure*} 

\vspace{5mm}
\noindent
\textbf{\large Discussion} 

\vspace{2mm}
\noindent
In the previous sections we showed that the transport dynamics 
of molecular motors depend on the initial conditions of motion. 
This can be more clearly seen in Figs.~\ref{Fig:5}(A,D), where 
the MSD for the motion along a single filament is plotted for 
a given set of $\kappa_w$ and $\kappa_m$ parameters and for 
different probabilities $P_0^M$ of initially starting in the 
motion state (see also Figs.~\ref{Fig:5}(B,E) for the influence 
of initial conditions on the anomalous exponent $\alpha$). 
Indeed, the probabilities of finding the walker in each of the 
two states of motility at time $n$, i.e.\ $P_n^M$ and $P_n^W{=}
1{-}P_n^M$, are controlled by the transition probabilities $\kappa_w$ 
and $\kappa_m$ at long times. Thus, the influence of initial 
conditions gradually weakens as $P_n^M$ and $P_n^W$ converge 
towards their stationary values. The sequence of motion and 
waiting states can be considered as a discrete time Markov 
chain with transition probabilities $\kappa_m$ and $\kappa_w$. 
Denoting the initial probabilities by $P_0^M$ and $P_0^W{=}
1{-}P_0^M$, it can be verified that the time evolution of 
these probabilities follows
\begin{align}
P_n^M &= \frac{\kappa_m}{\kappa_m+\kappa_w} 
+ \frac{\left(1-\kappa_m-\kappa_w\right)^n}{\kappa_m
+\kappa_w}\left(\kappa_w P_0^M-\kappa_m(1
-P_0^M)\right), \notag \\
P_n^W &= \frac{\kappa_w}{\kappa_m+\kappa_w} 
- \frac{\left(1-\kappa_m-\kappa_w\right)^n}{
\kappa_m+\kappa_w}\left(\kappa_w P_0^M-\kappa_m(1
-P_0^M)\right). 
\label{Eq:MarkovP}
\end{align}
The probabilities eventually converges to the stationary values 
\begin{align}
P^M_{\!\infty}{=}\frac{\kappa_m}{(\kappa_m{+}\kappa_w)},\notag  \\
P^W_{\!\infty}{=}\frac{\kappa_w}{(\kappa_m{+}\kappa_w)},
\label{Eq:MarkovPstationary}
\end{align}
thus, if the process is initially started with these 
probabilities, the system is already equilibrated. 
Otherwise, the dynamics at short times is influenced 
by the choice of initial conditions of motion until 
the system relaxes towards stationary state, as shown 
in Figs.~\ref{Fig:5}(C,F). 

\begin{figure}[b]
\begin{minipage}{0.5\textwidth}\hspace*{-7pt}
\includegraphics[width=\textwidth]{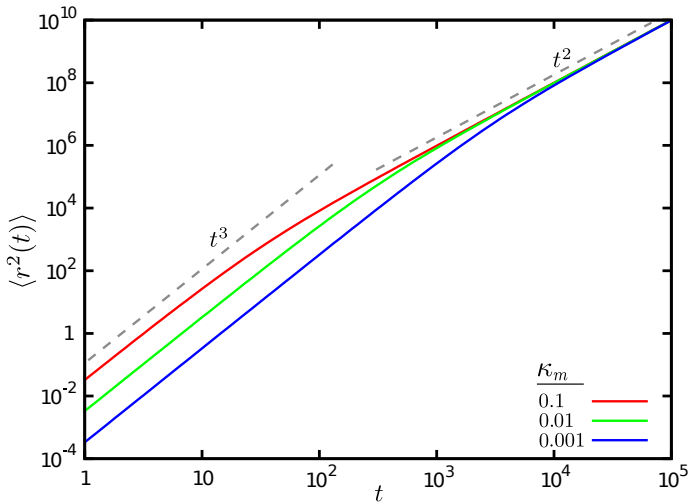}
\end{minipage}\hfill
\begin{minipage}{0.35\textwidth}
\caption{MSD for the motion on a single filament, starting 
from the initial condition $P_0^M{=}0$ and without any switching 
from the waiting to motion state, i.e.\ $\kappa_w{=}0$. The 
results are shown for $\lambda{=}1$ and several values of 
$\kappa_m$.}
\label{Fig:6}
\end{minipage}\hspace*{-2.5pt}
\end{figure}

\smallskip\smallskip\smallskip
\noindent
By varying the probability $P_0^M$ of initially starting 
in the motion state, we find regimes with an anomalous 
exponent greater than $2$ in Figs.~\ref{Fig:5}(B,E). To 
understand the origin of this peculiar behavior we note 
that the presented MSD is indeed an ensemble averaged 
quantity. When starting from the initial probability 
$P_0^M$ which is smaller (larger) than the equilibrated 
value $P^M_{\!\infty}$, an acceleration (deceleration) 
due to the injection of more (less) particles from the 
waiting to the motion state occurs. This is especially 
more obvious for the case of motion along a single 
filament with $P_0^M{=}\kappa_w{=}0$. In this case, 
all particle are in the waiting state initially, 
and no transition to the waiting state happens during 
the process, thus, all particles switch to the motion 
state gradually according to the transition probability 
$\kappa_m$. By injection of more particles into 
the ballistic motion state, the exponents greater than 
$2$ appear, as shown in Fig.~\ref{Fig:6} (see also the 
supplementary figure S2). The exponent converges to $2$ 
as the system approaches the stationary state. 
Figure~\ref{Fig:6} shows that the initial slope of the 
ensemble averaged MSD is nearly $3$ for this set 
of the parameter values, which can be explained as 
follows. Because of the constant transition probabilities, 
the residence time in the waiting state is exponentially 
distributed. The particles of the ensemble, which are 
all in the waiting phase initially, switch to motion 
at exponentially distributed times $t_0$, i.e.\ $p(t_0)
{=}\kappa_m e^{-\kappa_m t_0}$. As soon as a particle 
switches the state, it performs a ballistic motion with 
velocity $v$, thus 
\begin{align}
r(t)=
\begin{cases} 0, &t<t_0 \\ v\left|t-t_0\right|,
&t\geqslant t_0 \end{cases}
\end{align}
and we obtain the ensemble averaged MSD by integration over 
all possible transition times until time $t$
\begin{align}
\langle r^2(t) \rangle = \int_0^t \text{d}t_0~p(t_0)
v^2\left|t-t_0\right|^2 = \frac{v^2}{\kappa_m^2}
\left[2-2\kappa_mt+\kappa_m^2t^2-2e^{-\kappa_mt}\right].
\end{align}
By Taylor expansion around $t{=}0$, one finds $\langle 
r^2(t) \rangle \simeq \frac{v^2}{\kappa_m^2}
\left[\frac{1}{3}\kappa_m^3t^3+\dots\right]$, 
which verifies that the initial anomalous exponent 
is nearly $3$. A similar procedure can be applied to 
obtain the initial exponents for other sets of 
parameter values, as those shown in Figs.~\ref{Fig:5}(B,E).

\smallskip\smallskip\smallskip
\noindent
In summary, we developed a general analytical framework to 
study the dynamics of molecular motors on cytoskeleton, which 
enabled us to identify the influence of filament network 
structure, moving/pausing probabilities, and motor processivity 
on the transport properties of motors. The flexibility of 
our coarse-grained formalism allowed us to consider different 
filamentous structures, from a single filament to a complex 
network of biopolymers characterized by its structural 
heterogeneity and anisotropy. We obtained exact analytical 
expressions for the arbitrary moments of displacement, and 
verified that the motors display a wide range of different 
types of motion due to the interplay between motor processivity, 
structural properties of filamentous network, and transition 
probabilities between the two states of motility. One observes that 
multiple crossovers occur between different types of anomalous 
transport at short and intermediate timescales, and that 
the crossover time to the asymptotic diffusive or ballistic 
motion varies by several orders of magnitude. We also addressed 
how the initial conditions of motion affect the resulting 
dynamics.

\smallskip\smallskip\smallskip
\noindent
In the analysis presented in this work, the transitions between 
the two states of motion were assumed to be spontaneous, thus, the 
motion of motors was described by a Markovian process. However, 
switching between the states might be not necessarily spontaneous, 
and the process can be non-Markovian in general, e.g.\ if the 
residence time in a state affects its switching probability to 
the other state. These generalizations can be handled within our 
proposed analytical framework enabling to deal with different 
types of environments. The analytical formalism is also applicable 
to study similar dynamics in other biological as well as nonliving 
systems, such as the run-and-tumble motion of bacteria in biological 
media. It is worthwhile to mention that the motility in the passive 
state may not be negligible in general. The dynamics in such a case 
can be described by altering states of motion with different 
velocities and run times.

\vspace{5mm}
\noindent
\textbf{\large Methods} 

\vspace{2mm}
\noindent
We develop an analytical Fourier-z-transform formalism to describe 
the persistent motion of motor proteins along cytoskeletal filaments 
with stochastic pausing periods. A random walk in discrete time and 
continuous space with two states of motility is introduced. The walker 
either performs a persistent motion on filaments or waits when faces 
obstacles or tumbles in the crowded cytoplasm. By introducing the 
probability density functions $P_{n}^{M}(x,y|\theta)$ and $P_{n}^{W}(
x,y|\theta)$ for the probability to find the walker at position $(x,y)$ 
along the direction $\theta$ at time step $n$ in the motion or waiting 
states, the temporal evolution of the process can be described by a 
set of coupled master equations, where the transitions from motion 
to waiting state and vice versa (denoted by $\kappa_w$ and $\kappa_m$, 
respectively) are assumed to be asymmetric and constant. Then, arbitrary 
moments of displacement can be calculated by applying Fourier and z 
transforms on the set of master equations. Finally, the quantities 
of interest can be obtained in the real space and time by means of 
inverse transforms. A detailed description of the analytical approach 
can be found in \emph{Suppl Info}.

\bigskip

\noindent
\textbf{\sffamily Acknowledgments} 

\noindent 
This work was funded by the Deutsche Forschungsgemeinschaft (DFG) 
through Collaborative Research Center SFB 1027 (Projects A3 and A7).
\vspace{3mm}

\noindent{\textbf{\sffamily Author~contributions} 

\noindent 
MRS and LS designed the research; AEH and MRS developed the analytical 
framework; AEH performed simulations; All authors contributed to the 
analysis and interpretation of the results. AEH and MRS drafted the 
manuscript.
\vspace{3mm}

\noindent{\textbf{\sffamily Additional information} 

\noindent 
\textbf{Supplementary information} accompanies this paper at 
\href{http:/}{http://www.nature.com/srep}. 

\noindent 
\textbf{Competing financial interests:} The authors 
declare no competing financial interests.}
}

\end{document}